\documentclass[12pt]{article}
\usepackage{epsfig}
\def\be{\begin{equation}}
\def\ee{\end{equation}}
\def\bc{\begin{center}}
\def\ec{\end{center}}
\def\bea{\begin{eqnarray}}
\def\eea{\end{eqnarray}}

\catcode`@=11
\def\marginnote#1{}
\newcount\hour
\newcount\minute
\newtoks\amorpm
\hour=\time\divide\hour by60
\minute=\time{\multiply\hour by60 \global\advance\minute by-\hour}
\edef\standardtime{{\ifnum\hour<12 \global\amorpm={am}%
        \else\global\amorpm={pm}\advance\hour by-12 \fi
        \ifnum\hour=0 \hour=12 \fi
        \number\hour:\ifnum\minute<10 0\fi\number\minute\the\amorpm}}
\edef\militarytime{\number\hour:\ifnum\minute<10 0\fi\number\minute}
\def\draftlabel#1{{\@bsphack\if@filesw {\let\thepage\relax
   \xdef\@gtempa{\write\@auxout{\string
      \newlabel{#1}{{\@currentlabel}{\thepage}}}}}\@gtempa
   \if@nobreak \ifvmode\nobreak\fi\fi\fi\@esphack}
        \gdef\@eqnlabel{#1}}
\def\@eqnlabel{}
\def\@vacuum{}
\def\draftmarginnote#1{\marginpar{\raggedright\scriptsize\tt#1}}
\def\draft{\oddsidemargin 0.0truein
        \def\@oddfoot{\sl preliminary draft \hfil
        \rm\thepage\hfil\sl\today\quad\militarytime}
        \let\@evenfoot\@oddfoot \overfullrule 3pt
        \let\label=\draftlabel
        \let\marginnote=\draftmarginnote
   \def\@eqnnum{(\theequation)\rlap{\kern\marginparsep\tt\@eqnlabel}%
\global\let\@eqnlabel\@vacuum}  }
\catcode`@=12
\begin{document}
\begin{titlepage}
\vspace*{-1cm}

\begin{flushright}
hep-ph/0110234
\end{flushright}

\vskip 2.0cm

\begin{center}
{\Large\bf Supersymmetry and Brane Cosmology}
\end{center}
\vskip 1.0cm

\begin{center}
{\sc 
Antonio Riotto~$^{a,b,}$\footnote{E-mail: antonio.riotto@pd.infn.it}
and Luca Scarabello~$^{a,}$\footnote{E-mail: luca.scarabello@pd.infn.it}
}\\
\vskip .9cm
$^{a}$~{\it  University of Padova, 
Department of Physics}\\
\vskip .2cm
{\it  Via Marzolo~8, I-35131 Padova, Italy}\\
\vskip .5cm
$^{b}$~{\it INFN, Sezione di Padova,  Via Marzolo~8, I-35131 Padova, Italy}
\end{center}

\vskip 1.5cm

\begin{abstract}
\noindent
We consider a five-dimensional brane world scenario
where the fifth dimension is compactified on
$S^1/Z_2$. We show that the 
familiar four-dimensional cosmology on our brane is easily  
recovered during a primordial stage of inflation 
if   supersymmetry is exploited.
Even if  some  vacuum energy density  
appears localized on our  three brane, heavy supersymmetric bulk fields  
adjust themselves and acquire a nontrivial 
configuration along the extra-dimension. This phenomenon 
redistributes uniformly the    energy density across the bulk and the
resulting  energy-momentum tensor does not
display any   singularity associated to the initial localized  energy
density on our  three-brane. No jumps across the brane
are present for the derivatives of the metric and 
Einstein's equations are solved by 
constant solutions along the fifth dimension. Our findings make it
clear that  cosmological 
phenomena in the supersymmetric brane world scenario  must be 
studied taking properly into account bulk supersymmetric states. 
This comment is particularly relevant when applied to  (super)gravity since
in supersymmetric 
brane world scenarios, even though 
chiral matter and gauge fields may be  restricted 
to live on  boundaries, gravity  multiplets 
always propagate in the
bulk.
\end{abstract}

\end{titlepage}
\setcounter{footnote}{0}
\vskip2truecm

\section{Introduction}

The goal of this paper is to investigate the impact of supersymmetry
on the cosmology of the brane world.
The recent exciting developments in string 
theory and the idea that our Universe may be thought
as a three-brane embedded in a higher dimensional theory has lately 
stimulated a lot
of activity in various fields of research. In the cosmological setting, 
it has been shown that 
a non-standard cosmological evolution of our Universe is induced if  
matter with
energy density $\rho$ is confined on 
three branes \cite{b1,b2}. The Friedmann equation governing the rate of 
the expansion of our 
three-brane Universe $H_0$
is modified and one finds $H_0^2\propto \rho^2$, instead of 
the conventional four-dimensional 
Hubble law $H_0^2\propto \rho$. This result is essentially due to nontrivial
constraints on the derivatives of the scale factor along the extra-dimensions 
and on
the energy densities when the latter are localized on the branes.
A strong constraint on the brane world idea would  then be  provided
by the requirement of having a  standard cosmological 
evolution which successfully describes
our Universe from the epoch of 
nucleosynthesis to the present day.

An elegant solution to this problem is offered within the Randall-Sundrum 
setting 
\cite{rs1,rs2}
where the tension of the brane is 
compensated by a negative cosmological constant in the bulk. 
The standard Hubble law is almost recovered
if the energy density on the brane is 
much smaller than the brane tension \cite{cs}. Indeed, in the
phenomenologically interesting model which solves the 
hierarchy problem (in which our Universe
is identified with a three-brane with negative tension), there is a crucial
sign difference in  the Friedmann equation. This obstacle is futher
 overcome if one  
takes into account
that the so-called radion,  the four-dimensional
modulus parametrizing the radius of compactification, 
has to be stabilized \cite{terning,kogan}.
This clarifies that the  origin of the
unconventional cosmology is not due to the breakdown of the effective 
four-dimensional
theory, but rather to a constraint that matter on 
the branes is forced to obey in order to ensure
a static radion modulus. Upon  radion stabilization, solutions can be found
for  in 5D for 
 the $3$-space scale factor $a(t,y)$ which have a   nontrivial dependence 
on the
coordinate $y$ of the extra-dimension  and  a local minimum at some point 
$y=y_m$. If the
theory  is compactified on a circle with radius smaller than
$|y_m|$,  normal Friedmann expansion is 
obtained.

In  this paper we will  show that the 
familiar four-dimensional Hubble law in the brane world scenario can be 
recovered
-- at least
during a primordial stage of inflation -- if one exploits  supersymmetry.
This result has a simple explanation. Suppose, for instance, 
that  one starts from a gauge
theory  extended to the
entire bulk. If supersymmetry is imposed, the theory necessarily comes
with  heavy scalar fields which are contained in  the supersymmetric multiplets
and do not have a massless zero mode.
They are
coupled to the fields on the three-brane
only through
derivatives along the extra-dimension coordinates. 

If some  energy density  
appears localized on our 
three brane, these fields  adjust themselves and acquire a nontrivial 
configuration
along the extra-dimensions. This back-reaction 
redistributes uniformly the    energy density across the bulk. In
 the effective four-dimensional theory the energy-momentum tensor does not
display any singularity which would signal a localized source on our 
three-brane.
Einstein's
equations admit constant solutions for the scale factor across the bulk and 
the familiar four-dimensional Hubble law may be recovered. 

Admittedly, this gratifying result has been found only for the specific
epoch of the evolution of the Universe when the latter undergoes
a period of accelerated expansion.
Nevertheless, an important lesson can be learned from our simple
exercise. When 
investigating various phenomena occured in the early Universe within the
brane world scenario -- inflation,
reheating after inflation, phase transitions, generation of the baryon 
asymmetry, etc. -- a careful treatment is needed to properly  take into 
account 
all the degrees of freedom of the theory.
 Even those  states which
are massive and might seem irrelevant in the 4D effective description
may play a crucial role and significantly alter 
(and possibly simplifying) the description of 
the cosmological evolution. This remark holds especially for those states
living in the  (super)gravity multiplets which are necessarily present
in any supersymmetric construction of the brane world.
We will come back to this point  
in the last  section and discuss explicit examples.

The paper is organized as follows. In section 2 we consider 
a simple supersymmetric model of inflation whose dynamics 
is entirely confined on a boundary
wall and 
briefly summarize the findings of Refs. \cite{terning,kogan} to show how the 
standard 4D cosmological evolution may be obtained. In section 3 we present
our results staring from a theory where supersymmetry is extended to the 
whole
bulk. Section 4 contains our conclusions and a  discussion of the 
implications of
our findings.

\section{Conventional cosmology in the brane world}

Our starting point is a five-dimensional theory compactified 
on an orbifold $S^1/{Z}_2$ of (comoving) radius $R$. One writes the Lagrangian as
\begin{equation}
  S =   \frac{1}{2}\int\, d^4x\,\int_{-\pi R}^{+\pi R}\,dy \,
\sqrt{g_5}\,
\left\{ {\cal L}_{\rm bulk}
        + \sum_i \delta(y- y_i){\cal L}_{4i} \right\}\ .
\label{totalS}\end{equation}
The sum includes the  walls at the orbifold points 
$y_i = 0, \pi R$. The bulk
Lagrangian ${\cal L}_{\rm bulk}$ includes  the standard 5D gravity Lagrangian
\begin{equation}
{\cal L}_{\rm bulk}=-\frac{1}{2} M_5^3 R_5\, ,
\end{equation}
where $M_5$ is the five-dimensional reduced Planck mass and $R_5$ is the 
five-dimensional scalar curvature.
The five-dimensional metric is written as by 
\begin{equation}
ds^2 = g_{AB} dx^A dx^B=g_{\mu\nu} dx^\mu dx^\nu-b^2(t) dy^2
= n^2(y,t) dt^2 - a^2(y,t) d \vec{x}^2 - b^2(t) dy^2\, ,
\label{metric}
\end{equation}
where the five-dimensional coordinates are indicated by $x^M = (x^\mu,y)$
and $g_{\mu\nu}$ denotes the usual four-dimensional metric on hypersurfaces
of fixed  $y$. 
The latter 
parametrizes the extra dimension compactified on 
the interval $\left[ -\pi R,+\pi R \right]$ and the $Z_2$ 
symmetry $y \leftrightarrow - y$ is imposed. Our four-dimensional brane world is supposed to
be at $y=0$.

Under the aforementioned decomposition (\ref{metric})
and after a conformal transformation
$g_{\mu\nu} \rightarrow b^{-1} g_{\mu\nu}$,
 the action 
(\ref{totalS}) can 
be written as 
\begin{equation}
S=\frac{1}{2}\int\, d^4x\,\int_{-\pi R}^{+\pi R}\,dy \,
\sqrt{-g_4}
\,\left\{-\frac{1}{2} M_5^3\,
\left[R_4(t,y) - \frac{3}{2}\, \dot{r}^2\right]
+ e^{-r}\,{\cal L}_{\rm st}\right\}\, ,
\label{s1}
\end{equation}
where  $R_4$ is the four dimensional scalar curvature and 
we have explicitly inserted a Lagrangian ${\cal L}_{\rm st}$
which is resposible for the stabilization of the 
 radion field $r\equiv {\rm ln}\, b$.

Einstein's equations  are given by (after  radion stabilization)
\begin{eqnarray}
G_{00} &=& 3 \left( \frac{\dot{a}}{a} \right)^2
- 3 \frac{n^2}{b^2} \left[ \frac{a''}{a}
+ \left( \frac{a'}{a} \right)^2 \right]=\frac{T_{00}}{M_5^3} \, ,
\label{00} \\
G_{ii} &=& \frac{a^2}{n^2} \left[ -  \left( \frac{\dot{a}}{a} \right)^2
+ 2 \frac{\dot{a}}{a} \frac{\dot{n}}{n} - 2 \frac{\ddot{a}}{a} \right]
+ \frac{a^2}{b^2}
\left[ \left( \frac{a'}{a} \right)^2 + 2 \frac{a'}{a} \frac{n'}{n}
+ 2 \frac{a''}{a} + \frac{n''}{n} \right]=\frac{T_{ii}}{M_5^3}\, , 
\label{ii}\\
G_{55} &=& 3 \left[ \frac{a'}{a}\left( \frac{a'}{a}+\frac{n'}{n}\right)
-\frac{b^2}{n^2}\left(\frac{\dot{a}}{a}\left(\frac{\dot{a}}{a}-\frac{\dot{n}}{n}
\right)+\frac{\ddot{a}}{a}\right)\right]=\frac{T_{55}}{M_5^3}\, , 
\label{55}\\
G_{05} &=& 3 \left[ \frac{n'}{n} \frac{\dot{a}}{a} - 
\frac{\dot{a}'}{a} \right]=\frac{T_{05}}{M_5^3}
\label{G}\, ,\label{05}
\end{eqnarray}
where dot denotes differentiation whith respect to $t$, prime 
with respect to $y$ and $T_{MN}$ is the energy-momentum
tensor.

\subsection{Inflation driven by a boundary vacuum energy}

Since we are interested in the case in which our  Universe goes 
 through an inflationary stage, we first assume that 
there is a nonvanishing vacuum energy $V$ on our  brane at $y=0$. The 
corresponding energy-momentum tensor 
can be expressed in the form
\begin{equation}
\left.T^A_{\,\,\,B}\right|_{\rm brane}= \frac{\delta (y)}{b} \, {\rm diag}
\left(V,V,V,V,0 \right)\, .
\label{source}
\end{equation}
To be concrete, we suppose that inflation is driven by a nonvanishing 
supersymmetric
$D$-term  \cite{dterm} (the same considerations hold for 
$F$-term inflation). 
To exemplify the description,
let us consider an abelian $U(1)$ gauge theory  on our brane
(therefore gauge fields do not propagate in the bulk) with
coupling constant $g$. The theory  
contains  three chiral superfields
on the boundary at $y=0$:  
$S$,
$\Phi_+$ and
$\Phi_-$ with charges equal to $0$, $+ 1$ and $- 1$
respectively under the $U(1)$ gauge symmetry.
The superpotential on the boundary has the form
\begin{equation}
W = \lambda S\Phi_+\Phi_- \, 
\end{equation}
and the Lagrangian contains the Fayet-Iliopoulos $D$-term
\begin{equation}
{\cal L}_{\rm FI}=D\, \xi\, .
\label{dterm}
\end{equation}
The scalar potential in the global supersymmetry limit reads
\begin{equation}
V = \lambda^2 |S|^2 \left(|\phi_-|^2 + |\phi_+|^2 \right) +
\lambda^2|\phi_+\phi_-|^2 +
{g^2 \over 2} \left(|\phi_+|^2 - |\phi_-|^2  + \xi \right)^2
\end{equation}
where $\phi_{\pm}$ are the scalar fields of the supermultplets
$\Phi_{\pm}$.

The global minimum is supersymmetry conserving, but the gauge group
$U(1)$ is spontaneously broken
\begin{equation}
\langle S \rangle  = \langle \phi_+  \rangle = 0, ~~~ \langle \phi_-\rangle
= \sqrt{\xi}\, .
\end{equation}
However, if we minimize the potential, for  fixed values of $S$,
with respect to
other fields, we find that for  $S > S_c = {g \over \lambda}
\sqrt{\xi}$, the minimum is at $\phi_+ =\phi_- = 0$. Thus, for
$S > S_c$ and $\phi_+ =\phi_- = 0$ the tree level potential
has a vanishing curvature in the $S$ direction and large positive
curvature in the remaining two directions $m_{\pm}^2 = 
\lambda^2|S|^2 \pm g^2\xi$. 
For arbitrarily large $S$ the vacuum energy density driving inflation
is provided by the tree level value of the potential 
\begin{equation}
V = {g^2 \over 2}\xi^2
\label{vacuum}
\end{equation}
and the
$S$ plays the role of the   inflaton. Notice that under these circumstances
the $D$-term $D=\xi+|\phi_+|^2 - |\phi_-|^2$ reduces to 
\begin{equation}
D=\xi\, .
\label{xi}
\end{equation}
One-loop corrections generate an almost flat potential for the 
inflaton field $S$ and 
the end of inflation is determined 
either by
the failure of the slow-roll conditions or when $S$ approaches $S_c$ 
\cite{dterm}.

Since there is no flow of matter
along the fifth  dimension both $T_{05}$ and  $G_{05}$ vanish. 
The $(0,5)$-component   of Einstein's  equations can be easily 
integrated to give
\begin{equation}
n(t,y)=\lambda(t)\,\dot{a}(t,y)\, .
\end{equation}
The
$(0,0)$-component   of Einstein's  equations
reduces to a second-order differential equation for $a(t,y)$ while the function
$\lambda(t)$ leads not only to the determination of 
the lapse function $n(t,y)$, but also
to the four-dimensional Friedmann equation on our  brane where the Hubble 
parameter
can be expressed in terms of $\lambda(t)$ as
\begin{equation}
H^2_0\equiv\left(\frac{\dot{a}_0}{a_0}\right)^2
=\frac{1}{\lambda^2(t)a_0^2(t)}
\, .
\label{hubble}
\end{equation}

The brane can  be  taken into account by using the junction
conditions  which relate the jumps of the derivative of the metric across
 the brane to the stress-energy tensor (\ref{source}) inside the brane
\cite{b1,b2}. This gives
\begin{equation}
\frac{[a^\prime]}{a_0 b_0}=-\frac{1}{3}\frac{V}{M_5^3}\, ,
 \label{aarho}
\end{equation}
where the subscript $0$ for  $a_0$ and $b_0$ means that these functions are
taken in $y=0$, and $[f]= f(0^+)-f(0^-)$
denotes the jump of the function $f$ across $y = 0$.

The general solution for $a(t,y)$ can be written as 
\begin{equation}
a^2(t,y)=a_0^2(t)+\alpha(t)\,|y|+\frac{b_0^2}{\lambda^2}\, y^2\, ,
\label{4d}
\end{equation}
with
\begin{equation}
\alpha(t)=-\frac{a_0^2\, b_0}{3}\frac{V}{M_5^3}\, .
\end{equation}
Notice that $a^2(t,y)$ has a minimum at $|y_m|(t)=
-\alpha(t)\lambda^2(t)/2b_0^2$, which forces 
to compactify the extra dimension on a circle of radius $|y_m|$
by identifying the two extrema at $-|y_m|$ and $|y_m|$.

From Eq. (\ref{hubble}) one derives \cite{kogan}
\begin{equation}
H_0^2=\frac{1}{\lambda^2(t)a_0^2(t)}=\frac{V}{6\,b\,|y_m|\,M_5^3}\, 
\end{equation}
from which it is concluded that one can recover the conventional 4D 
Friedmann equation
only if $|y_m(t)|=$ constant. In such a case, one can identify the 
four-dimensional reduced
Planck mass 
\begin{equation}
M_4^2=2\,b\,|y_m|\,M_5^3\, . 
\end{equation}

Let us suppose that 
a bulk potential for the radion field $V \left( b \right)$ is generated in
the five-dimensional theory by some mechanism and  the radion is
very heavy, that is if near the minimum $b_0$ we have $V \left(
b \right) \simeq M_5^5 \left( b - b_0 \right)^2 / b_0^2$ with a
very high mass scale $M_5$.
Since $
\left(T^\mu_{\,\,\,\mu}-2\, T^5_{\,\,\, 5}\right)$ is the source for the
radion modulus, the latter 
 remains in equilibrium if the energy momentum tensor 
satisfies
the following constraint \cite{kogan}
\begin{equation}
\int_{-|y_m|}^{+|y_m|}\,dy \,\sqrt{-g_4}\, e^{-r}\, 
\left(T^\mu_{\,\,\,\mu}-2\, T^5_{\,\,\, 5}\right)=0\, .
\label{constraint}
\end{equation}
Since during the inflationary stage on our brane there is an extra vacuum 
energy density 
(\ref{vacuum}), the equilibrium position for the radion field changes. 
Using the
constraint (\ref{constraint}) (with the integration 
over the fifth coordinate 
$y$
now going from $-|y_m|$ to $|y_m|$) and 
 $\left.T^\mu_{\,\,\,\mu}\right|_{\rm brane}=4\,\frac{\delta(y)}{b}\, V$, 
one finds 
that
the minimum of the radion field is shifted by a small amount if $V$ is much 
smaller than 
$M_5^4$ and that the $(55)$-component of the energy momentum tensor becomes 
(up to order
${\cal O}(V^2)$)
\begin{equation}
T^5_{\,\,\,5}=\frac{V}{b_0\,|y_m|}\, .
\end{equation}
Under these circummstances $y_m(t)$=constant, 
as 
one can easily check plugging the solution
(\ref{4d}) into the $(55)$-component of Einstein's equations \cite{kogan}, and
the Friedmann equation for our Universe becomes 
\begin{equation}
H_0^2=\frac{1}{3}\,\frac{V}{M_4^2}\, .
\end{equation}
The conventional four-dimensional Hubble rate is recovered on our
three-brane at the expense of  limiting the
space available in the extra-dimension and compactifying on a circle
of radius $|y_m|\leq \pi R$.

Our goal is now to show that the conventional four-dimensional
cosmology (at least during the inflationary stage) 
is recovered when making use of all the tools offered by  supersymmetry.
 As we will show, our path towards standard 4D cosmology
differs considerably from the one outlined in this section.

\section{Supersymmetry and conventional four-dimensional cosmology}

We consider a simple variant of the model  of inflation discussed in the previous
section and suppose that the abelian gauge theory $U(1)$ lives in the bulk.
Gauge fields are therefore free to propagate in the extra dimension.

The five-dimensional $U(1)$  gauge multiplet with coupling
constant $g$ contains a vector field $A^M$, a real scalar field
$\Phi$, and a gaugino $\lambda^i$. The five-dimensional Yang-Mills 
multiplet is then extended to an
off-shell multiplet by adding an $SU(2)$ triplet $X^a$ of real-valued
auxiliary fields. Here capitalized indices
$M,N$ run over 0,1,2,3,5, lower-case indices $\mu$ run over 0,1,2,3, and 
$i$, $a$ are internal $SU(2)$ spinor and vector indices,
 with $i = 1,2$, $a= 1,2,3$.

Now we have to  project this structure down to a four dimensional $N=1$
supersymmetry transformation acting on fields on the orbifold points. A
generic bulk field $f(x^\mu,y)$ transforms under the action of the
$Z_2$-symmetry as  $f(x^\mu,y) = P\, f(x^\mu,-y)$
where $P$ is an intrinsic parity equal to $\pm 1$.  The quantum number $P$
must be assigned to fields in such a way that it leaves the bulk Lagrangian
invariant.  Then fields of $P= -1$ vanish on the walls but have nonvanishing
derivatives $\partial_5 f$.  

We assign even $Z_2$--parity to the fields
\be
A^\mu,\,\, \lambda_L^1,\,\, X^3\, ,
\ee 
and odd $Z_2$--parity to the fields
\be
A^5,\,\,\Phi,\,\, \lambda_L^2,\,\, X^1,\,\, X^2~\, .
\ee
On the wall at $y = 0$, the five-dimensional supersymmetry transformations 
 reduce to the following transformation of the
even-parity states generated by $\xi^1_L$:
\begin{eqnarray}
{\delta_{\xi}} A^{\mu}       & = 
	& i \xi^{1\dagger}_L \overline{\sigma}^\mu \lambda^1_L 
			- i   \lambda^{1\dagger}_L 
\overline{\sigma}^\mu \xi^1_L\, , 
			\nonumber \\ 
{\delta_{\xi}} \lambda^1_L & = 
	& \sigma^{\mu\nu}F_{\mu\nu}\xi^1_L -i 
(X^3 -\sqrt{-g^{55}} \partial_5 \Phi) 
			\xi^1_L\, , \nonumber \\ 
{\delta_{\xi}} X^{3}       & = 
	&  \xi^{1\dagger}_L \overline{\sigma}^\mu D_\mu \lambda^1_L
			+ \xi^{1\dagger}_L \sigma^2  
\sqrt{-g^{55}}\partial_5 \lambda^{2*}_L
			+ \, {\rm h.c.}\, , \nonumber \\ 
{\delta_{\xi}} \sqrt{-g^{55}}\partial_5 \Phi    & = 
	&  \xi_L^{1T} \sigma^2 \sqrt{-g^{55}} \partial_5 \lambda_L^2 +
 \xi_L^{1 \dagger} \sigma^2 \sqrt{-g^{55}} \partial_5 \lambda_L^{2*} \, .
\label{fourdtrans}
\end{eqnarray}
The last two equations imply
\begin{equation}
{\delta_{\xi}} (X^{3}-\sqrt{-g^{55}}\partial_5\Phi) 
	=  \xi^{1\dagger}_L \overline{\sigma}^\mu D_\mu \lambda^1_L
			+ \,  {\rm h.c.}\,    .
\label{Dtrans}
\end{equation}
A simple inspection of these transformations reveals that
the even  fields $A^\mu$, $\lambda^1_L$, 
and $(X^3 - \sqrt{-g^{55}}\partial_5\Phi)$ 
transform as the vector, gaugino, and 
the auxiliary $D$-field of a 4D $N=1$ vector multiplet \cite{mirpes}.

It is then obvious  how to couple the five-dimensional gauge multiplet to 
a generic 4D dimensional chiral multiplet 
living  on the boundary and charged under
the $U(1)$ symmetry \cite{mirpes}.
One writes the Lagrangian as 
\begin{equation} 
  S =   \frac{1}{2}\int\, d^4x\,\int_{-\pi R}^{+\pi R}\,dy \,
\sqrt{g_5}\,
\left\{ {\cal L}_{\rm gauge} 
        + \sum_i \delta(y- y_i){\cal L}_{4i} \right\}\ ,
\label{totalS1}
\end{equation}
where the sum includes the  walls at $y_i = 0, \pi R$. The bulk 
Lagrangian ${\cal L}_{\rm gauge}$ is  the standard one 
for a 5D super-Yang-Mills 
multiplet
\begin{eqnarray}
{\cal L}_{{\rm gauge}}   & = &{1\over g^2}\left( -\frac{1}{2}{\rm Tr}\,(F_{MN})^2 
	+ {\rm Tr}\,(D_M \Phi)^2
   + {\rm Tr}\,(\overline{\lambda} i  \gamma^M D_M \lambda)\right. 
\nonumber \\  
  &+&\left. {\rm Tr}\,(X^a)^2 - {\mbox{\rm Tr}}\,
	(\overline{ \lambda} [\Phi,\lambda])\right) \, ,
\label{fivedL}
\end{eqnarray}
with ${\rm Tr}\, [T^A T^B ] = \delta^{AB}/2$. 
The  boundary Lagrangian has the standard form of a four-dimensional
model built from the chiral multiplet charged under the $U(1)$ symmetry, but with a crucial difference:
the gauge fields $(A_\mu,\lambda_L,D)$ are replaced by the boundary values
of the bulk fields $(A_\mu, \lambda^1_L, X^3 - \sqrt{-g^{55}}\partial_5 \Phi)$.

\subsection{Inflation and conventional Hubble law}

In analogy with  Eqs. (\ref{dterm}) and
(\ref{xi}), we suppose that  the boundary chiral multiplets contain the   
fields $S$ and $\phi_{\pm}$ with the corresponding  
Fayet-Ilopoulos $D$-term on our brane contained in ${\cal L}_4$. 
The Fayet-Ilopoulos $D$-term
is now   written as
\begin{equation}
{\cal L}_{\rm FY}=\left(|\phi_+|^2 - |\phi_-|^2  + \xi \right)
\left(X^3 - \sqrt{-g^{55}}\partial_5 \Phi\right)\, .
\label{dterm1}
\end{equation}
Again, for very large values of inflaton $S$, the vacuum expectation values
of the fields $\phi_{\pm}$ are driven to zero and ${\cal L}_{\rm FY}$ reduces
to
\begin{equation}
{\cal L}_{\rm FY}=\xi\,
\left(X^3 - \sqrt{-g^{55}}\partial_5 \Phi\right)\, .
\label{dterm3}
\end{equation}
This $D$-term will be responsible for the inflationary stage.

With the action (\ref{totalS1}),
the boundary Fayet-Iliopoulos term (\ref{dterm1})
couples to the auxiliary field $X^3$
through the terms
\begin{equation}
  \frac{1}{2}\int\, d^4x\,\int_{-\pi R}^{+\pi R}\,dy \,
\sqrt{g_5}\,
\left\{{1\over g^2}{\rm Tr} 
	\,(X^3)^2  + \delta(y) \frac{\xi}{\sqrt{-g_{55}}}\,
   \left(X^3 - \sqrt{g^{55}}\partial_5 \Phi\right) \right\} \, .
\label{X}
\end{equation}
Integrating out the auxiliary field
$X^3$ through its equation of motion
\begin{equation}
X^3+g^2\,\frac{\xi}{\sqrt{-g_{55}}}\,\delta(y)=0 
\end{equation}
and including the kinetic term of the field $\Phi$, the singular terms
can be rearranged into a perfect square
\begin{equation}
\frac{1}{2\, g^2}\int\, d^4x\,\int_{-\pi R}^{+\pi R}\,dy \,
\sqrt{g_5}\,\left[\frac{1}{2}\,g^{\mu\nu}\partial_\mu\Phi\partial_\nu\Phi
+\frac{1}{2\,g_{55}}\left(\partial_5\Phi 
+g^2\,\xi \delta(y)\right)^2 \right]\, .
\label{five}
\end{equation}
At this point, it is worth emphasizing  that 
 the singular terms proportional to $\delta(y)$
and $\delta^2(y)$ play a crucial role at the quantum level since they 
 provide counterterms which are necessary in explicit computations
to preserve supersymmetry \cite{mirpes}. In particular, the role of the 
interaction term proportional to $\delta^2(y)$ is
to cancel the singular behaviour induced in diagrams where the
$\Phi$-field is exchanged.

From Eq. (\ref{five}) we can easily compute the energy momentum tensor
of the system
\begin{eqnarray}
\left.g^2\,T_{\mu\nu}\right|_\Phi&=& \partial_\mu\Phi\partial_\nu\Phi-g_{\mu\nu}\left[
\frac{1}{2}g^{\rho\sigma}\partial_\rho\Phi\partial_\sigma\Phi+
\frac{1}{2} g^{55}\left(\partial_5\Phi 
+g^2\,\xi \delta(y)\right)^2\right]\, ,\nonumber\\
\left.g^2\,T_{55}\right|_\Phi&=& \frac{1}{2}\left(\partial_5\Phi 
+g^2\,\xi \delta(y)\right)^2-\frac{1}{2}\,g_{55}\,g^{\rho\sigma}
\partial_\rho\Phi\partial_\sigma\Phi\, .
\label{em}
\end{eqnarray}
Notice the appearance of a potentially 
dangerous singular terms $\delta(y)$ and $\delta^2(y)$.

The following step amounts to integrating out the heavy field
$\Phi$. The reader should remember that, since  this field is odd under the
discrete $Z_2$-symmetry, it does not have a zero mode and all its modes are
as massive as the inverse of the radius of compactification.
This procedure makes therefore sense in the Kaluza-Klein approach whose 
purpose is to give a
four-dimensional interpretation of the five-dimensional world and  is supposed
to work when the  energy scale  of the system, in our case
the Hubble rate, is much smaller than the inverse of the radius
of the fifth dimension. 

Varying this action with respect to $\Phi$, we find that 
 $\Phi$ satisfies the equation
\begin{equation}
\partial_\mu\left(\sqrt{g_5}\, g^{\mu\nu}\,\partial_\nu\Phi\right)
+ \partial_5\left[\frac{\sqrt{g_5}}{g_{55}}
\left(\partial_5\Phi+g^2\,\xi \delta(y)\right)\right]=0\, .
\end{equation}

We now look for solutions such that $a$ and $n$ are independent of the
fifth coordinate $y$, such as $a(t,y)=a(t)$.
We can also  fix  $n(t)=1$ and suppose
that the radion is fixed at the minimum of its potential. 

Since $\Phi$ is an odd field under the $Z_2$-parity we have
$\Phi(0)=\Phi(\pi R)=0$ (where, for instance, $\Phi(0)$ has to be intended
as $(\Phi(0^+)+\Phi(0^-))/2$). 
For a static solution $\partial_\mu\Phi=0$, these 
boundary conditions of the field
$\Phi$ 
require that $\partial_5\Phi$ must integrate to zero around the circle
\begin{equation}
    \partial_5\Phi =
 - g^2 \xi \left( \delta(y) - {1\over 2\pi R}\right) \ .
\label{Phisol}\end{equation}
Substituting this solution into the Lagrangian (\ref{five}) one finds
that the various singular terms  cancel and one is left with
the usual $D$-term interaction
\begin{equation}
  S =  -\frac{1}{2}\int\, d^4x\,\int_{-\pi R}^{+\pi R}\,dy \,
\sqrt{g_4}\,\frac{1}{b_0}\,\frac{g^2}{2}\,\left(\frac{\xi}{2\pi R}\right)^2
=-
  \int\, d^4x\,\sqrt{g_4} \, \frac{\bar{g}^2 \xi^2}{4} 
\end{equation}
where $\bar{g}^2=(g^2/2\pi b_0 R)$. Correspondingly, the energy momentum
(\ref{em}) tensor reduces to
\begin{eqnarray}
\left.T_{\mu\nu}\right|_\Phi&=& \frac{g_{\mu\nu}}{\pi b_0 R}\, V\, ,
\nonumber\\ 
V &=&\frac{1}{4} \bar{g}^2 \xi^2\, .
\end{eqnarray}
Making use of the  Lagrangian (\ref{s1}) with
\begin{equation}
R_4=-6\left(\frac{\dot{a}^2}{a^2}+\frac{\ddot{a}}{a}\right) 
\end{equation}
or, equivalently, using the $(00)$-component of Einstein's equations,
we find that the conventional four-dimensional Hubble law   governing
the expansion rate of the Universe
\begin{equation}
H^2=\frac{1}{3}\frac{V}{M^2_4}\, ,
\end{equation}
where $M^2_4=\pi b_0 R M_5^3$ is the reduced four-dimensional Planck mass. 
One can also 
easily show that the $(55)$-component of Einstein's equations is satisfied 
once the
shift in the radion vacuum expectation value is taken into account.

This result is quite gratifying. The singular 
terms proportional to $\delta(y)$ and $\delta^2(y)$
disappear after
we substitute in the 5D Lagrangian the solution of the classical 
equation of motion for the heavy  field $\Phi$. The remarkable  
consequence is that the 
energy-momentum tensor is not peaked around the brane at $y=0$ and 
the constraint (\ref{aarho}) needs not to be imposed. Conventional 4D
 evolution is recovered.
The source of such findings  is manifest: 
supersymmetry imposes  the presence of the bulk 
propagating  field $\Phi$ in the effective auxiliary $D$-term on the
boundary. At the level of the effective 4D theory, all singular terms
disappear after we substitute in the Lagrangian the solution of the classical 
equation of motion of such odd field $\Phi$. Dynamically what happens is that
the bulk field $\Phi$ adjusts itself to response to any change in the $D$-term
on the boundary. This phenomenon is responsible for the 
redistribution of  the energy density in the bulk and for removing  the  
singular
terms at the boundary.

We close this section by reminding the reader that our finding hold as well 
in the
case in which the inflationary stage is driven by an $F$-term. In such a case,
 instead
of a vector multiplet in the bulk, one starts from a supersymmetric 5D 
hypermultiplet in the
bulk. The latter contains two scalar fields $A_1$ and $A_2$ which are even 
and odd respectively
under the $Z_2$ symmetry. The $F$-term on the boundary is 
\begin{equation}
F_1-\sqrt{-g^{55}} \partial_5 A_2
\end{equation}
leading to a boundary action 
\begin{equation}
{\cal L}_4= 
\left(F_1-\sqrt{-g^{55}}\partial_5 A_2\right)
\frac{\partial W}{\partial A_1}
\, .
\end{equation}
where $W$ is the boundary superpotential and $V_F=
\left|\frac{\partial W}{\partial A_1}\right|^2$
is the $F$-term vacuum energy density resposible for inflation.
Integrating out the auxiliary field $F_1$ one finds the bulk action
\begin{equation}
S=\frac{1}{2}\int\, d^4x\,\int_{-\pi R}^{+\pi R}\,dy \,
\sqrt{g_5}\,\left[g^{\mu\nu}\partial_\mu A_2^*\partial_\nu A_2+
\frac{1}{g_{55}}\left(\partial_5 A_2-\left|\frac{\partial W}{\partial A_1}
\right|\,
\delta(y)\right)^2
\right]\, .
\end{equation}
Again, one can  solve the equation of motion for the odd bulk field
$A_2$ and show that the singular terms proportional to $\delta(y)$ and 
$\delta^2(y)$
are cancelled leaving behind  an energy momentum tensor 
$\left.T_{\mu\nu}\right|_{A_{2}}= \frac{g_{\mu\nu}}{2\pi b_0 R}\, V_F$. 
The latter
guarantees the standard four-dimensional Hubble expansion during inflation.

\section{Conclusions and directions for future work}

We have shown that, if we start with a vacuum  energy density confined 
on our brane and the bulk is supersymmetric, 
the back-reaction of the bulk supersymmetric fields smooth out  the 
singularities
which would be otherwise displayed in the energy-momentum tensor of the system.
This considerably simplifies the solution of Einstein's equations since
derivatives of the metric do not jump across the brane, 
ensuring that 
uniform solutions along the fifth dimension can be found. 
Conventional 4D evolution
during the inflationary stage is recovered without resorting to any nontrivial
configuration of the scale factors along the extra-dimension and to any 
limitation
in the bulk, $|y|\leq |y_m|$.

Our results suggest that the study of cosmological 
phenomena in the brane world scenario  must be 
performed including all ingredients provided by supersymmetry. 
This comment is particularly relevant when applied to  gravity.
In supersymmetric 
brane world scenarios, even though 
chiral matter and gauge fields may be  restricted 
to live on the boundaries, gravity always propagates in the
bulk. Using an off-shell supergravity multiplet one can 
integrate out the auxiliary fields and examine the couplings between 
the on shell bulk supergravity fields and boundary matter fields 
\cite{zucker,kugo,us}.
The (super)gravity on shell multiplet contains
apart from the f\"unfbein 
$e_M^{\;\;A}$ and  the symplectic Majorana 
gravitino $\Psi_M$, the graviphoton $A_M$.
The situation is  analogous to what happens in the 
case of an off-shell bulk vector multiplet in 5D analyzed in section
3. There, the presence of the propagating odd field $\Phi$ in the 
effective $D$-term  $D=(X^3-\sqrt{-g^{55}}
\partial_5 \Phi)$ on the boundary induced 
new interactions between the the chiral matter fields living at the 
boundary and the $\Phi$ field. In supergravity 
one finds  new interaction
terms at the boundaries (compared to the usual ones in $N=1$ 4D supergravity
coupled to chiral or vector multiplets)
involving the components of the chiral and vector multiplets
and  the graviphoton field strength $F_{\mu5}$. The latter is
 made out of the
five-dimensional  field $A_\mu$ (odd under the $Z_2$ symmetry)  and the
four-dimensional field $A_5$ which  plays the role of the imaginary
part of the radion modulus.
If a chiral multiplet $(\varphi,\psi_\varphi,F_\varphi)$ 
lives on the boundary, the graviphoton couples to 
the  current
$J^\mu=i(\varphi^\dagger\partial^\mu\varphi-\varphi\partial^\mu
\varphi^\dagger)+
\psi_\varphi\sigma^\mu\bar{\psi}_\varphi$ 
forming  again a perfect square \cite{us}
\be
\label{neww}
S =   \frac{1}{4}\int\, d^4x\,\int_{-\pi R}^{+\pi R}\,dy \,
\sqrt{g_5}\,
  \left( \sqrt{-g^{55}}\,F_{\mu5} 
 - i \sqrt{\frac{3}{2}}\, J_\mu\,\delta(y)\right)^2~.
\ee
If during  the evolution of the  Universe the
current acquires a vacuum expectation value, one may not
disregard the axion-like coupling of the graviphoton field with the current.
On the contrary the latter acts as a source for the graviphoton. This 
happens, for instance, in the Affleck-Dine scenario
\cite{ad} where the generation of the baryon asymmetry is induced by 
time-dependent baryonic currents of scalar fields, or in the presence
of topological defects around which the imaginary part
of the scalar field the defect is made of  winds.

The same arguments tell us that the phenomenon of reheating after inflation
cannot be a purely four-dimensional event since the release of the vacuum 
energy density is accompanied by large fluctuations of the bulk fields.
For instance, in the $D$-term inflationary 
scenario,  the bulk quantity $\partial_5\Phi$
changes from $V^{1/2}$ during inflation to zero after inflation.
The same large fluctuations may occur during a primordial phase
transition. All these issues deserve further study and
are currently under investigation.

\section*{Acknowledgements}

This work was partially supported by the RTN European program ``Across
the Energy Frontier'', contract HPRN-CT-2000-00148.

\end{document}